# Opportunistic Content Search of Smartphone Photos

Ardalan Amiri Sani [*], Wolfgang Richter [§], Xuan Bao [†], Trevor Narayan [†],
Mahadev Satyanarayanan [§], Lin Zhong [*], Romit Roy Choudhury [†]
[*] Rice University, [§] Carnegie Mellon University, [†] Duke University

## ABSTRACT
Photos taken by smartphone users can accidentally contain content that is timely and valuable to others, often in real-time. We report the system design and evaluation of a distributed search system, Theia, for crowd-sourced real-time content search of smartphone photos. Because smartphones are resource-constrained, Theia incorporates two key innovations to control search cost and improve search efficiency. Incremental Search expands search scope incrementally and exploits user feedback. Partitioned Search leverages the cloud to reduce the energy consumption of search in smartphones. Through user studies, measurement studies, and field studies, we show that Theia reduces the cost per relevant photo by an average of 59%. It reduces the energy consumption of search by up to 55% and 81% compared to alternative strategies of executing entirely locally or entirely in the cloud. Search results from smartphones are obtained in seconds. Our experiments also suggest approaches to further improve these results.

## 1. Introduction
Modern smartphones allow us to take photos on the go, capturing whatever we find interesting. We do selectively share some of them with friends and even the public, e.g., through social network websites such as Facebook and Flickr. However, the majority of smartphone photos will not be shared, or possibly even transferred to another computer. Our work was motivated by many important scenarios in which photos captured by a smartphone user become vitally important to others, often in real-time. For example, when a child is lost during a holiday parade, photos by smartphone users nearby become very valuable to the police and parents [1]. As another example, a family photo may reveal a theft [2] (see Figure 1). As yet another example, a sports reporter would like to find the smartphone photos taken from the best angle at the time of a goal during a soccer game. The key question is: *How can an interested party find relevant smartphone photos, in real-time*? Relevance of a photo is not only determined by the metadata of the photo (e.g., time and location), but also by its content (e.g., "a girl with a red coat").

Our answer to this question is a distributed search service called *Theia*. Theia considers registered smartphones as distributed databases and allows a third party to compose a query and pushes it into these smartphones to find out photos that match the query. The query is a piece of code that examines not only the metadata

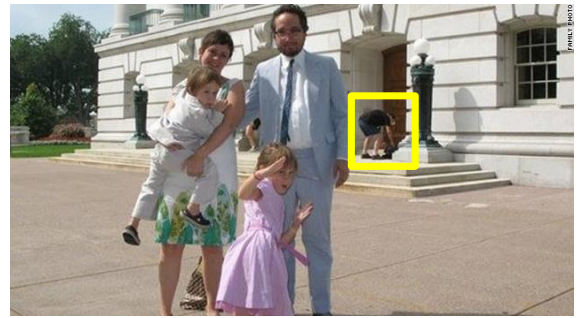

**Figure 1: Theft caught in the background of a family photo (Source: CNN [2]). Although this particular photo was not taken with a smartphone, it exemplifies the opportunistic value of photos taken by others**

but also the content of a photo. We focus on architecture and system design of Theia here, deferring issues such as incentive mechanisms and privacy control for future. In particular, we focus on *how Theia helps its users control search cost and improve search efficiency*. Unlike existing search systems whose databases are hosted by powerful data centers, Theia's databases are hosted by resource-constrained smartphones. Executing a query inside a smartphone can be resource-intensive and incur high cost to the smartphone owner that will eventually be paid by the search user. In view of the large number of smartphones Theia may search, the cost to the search user can be significant.

Theia incorporates two key innovations toward solving the above problem. *Incremental search* allows the search user to submit a cost budget along with a query and Theia will limit the search scope according to the budget. It tracks which photos have been searched by the query and allows the search user to effectively expand the scope by submitting the query again with a new budget. As in any search system, a search result, or a matched photo, is not necessarily what the search user is looking for or *relevant*. The objective of the incremental search is to help a search user find relevant photos with lowest cost per relevant photo. *Partitioned search* leverages the cloud to reduce the execution energy cost of a query in a smartphone. Based on the selectivity and energy cost of the predicates in the query and the wireless energy cost of offloading a photo, Theia dynamically identifies the predicates to be evaluated in the cloud and selectively offloads photos to reduce the energy cost of the smartphone.

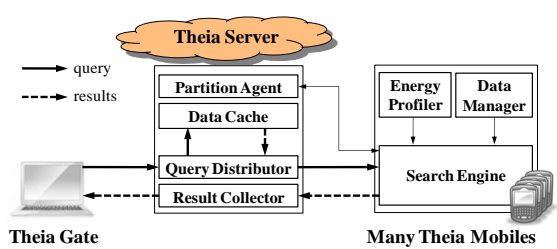

**Figure 2: Architecture of Theia and information flow between its components**

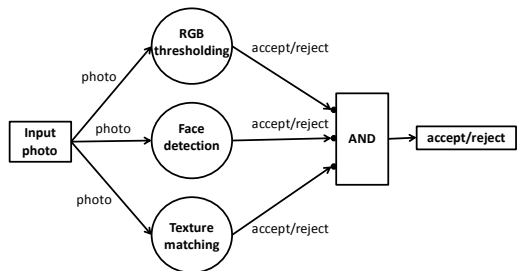

**Figure 3: An example Theia query (Query_1) that detects photos with a face and a large cloudy sky**

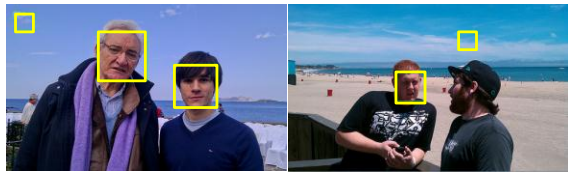

**Figure 4: Examples of smartphone photos accepted by Query_1 from Flickr**

We describe a complete, working prototype of Theia that consists of three components: Theia Server that distributes queries and runs in the cloud, Theia Mobile that executes queries and runs in registered smartphones, and Theia Gate, that allows a search user to compose and revise queries to examine the content of photos. Theia Server and Theia Mobile collaborate to implement incremental and partitioned search.

We report a three-part evaluation. First, a user study with 10 participants on a competitive search task spanning 85 emulated smartphones shows that Theia's incremental search reduces the cost per relevant photo by an average of 59%, and helps to retrieve 44% more relevant photos. Second, a measurement study demonstrates that Theia's partitioned search reduces the energy consumption of executing the search in the smartphone by up to 55% and 81% compared to alternative strategies of executing entirely locally or entirely in the cloud. The dynamic partition feature also enables Theia to adapt to changing network conditions. Finally, a field study with a testbed of 6 Android smartphones with photos from smartphones of real users show that Theia returns results with a median latency of seconds.

The rest of the paper is organized as follows. We will first present the Theia Architecture and then provide the key technical innovations of Theia, Incremental Search and Partitioned Search. We will present a full Prototype Implementation of Theia with Android smartphones. We offer the three-part Evaluation of Theia. We will also discuss Related Work before Conclusion.

## 2. Theia Architecture

As illustrated in Figure 2, Theia consists of three main components, *Theia Mobile* on smartphones that elect to participate, *Theia Server* on powerful servers in the cloud, and *Theia Gate* at the search user. Today, Theia Gate runs on laptops and desktops, but we anticipate creation of a smartphone implementation in the future. Using Theia Gate, a user generates a search query and submits it with a budget to Theia Server. Theia Server then distributes the query to selected smartphones according to the query's budget and execution history. At smartphones, Theia Mobile executes the query on the photos in the device and streams the photo results to Theia Server. Theia Gate streams results from Theia Server. The streaming aspect is important: the user starts seeing results even before query execution completes.

### 2.1 Theia Query

A Theia query is generated by the search user using Theia Gate. It takes a photo as input and outputs *accept* or *reject*. It is a logic combination (AND/OR) of *predicates*. A predicate takes a photo as input and outputs accept or reject, similar to the query itself.

A predicate is a piece of code that examines a specific feature in the content of the photo, e.g., people's faces, or specific metadata of the photo, e.g., time and location. Two important properties of a predicate are *selectivity* and *cost*. The selectivity of a predicate is the probability for photos to be accepted by it. The selectivity of two predicates may be correlated. This correlation can be quantified by the *conditional selectivity* of predicates. If $A_1$ and $A_2$ are the sets of photos accepted by predicates $p_1$ and $p_2$, respectively, the conditional selectivity $s(p1/p2)$ is $|A_1 \cap A_2| / |A_2|$. It is the probability that $p_2$ accepts the photos that $p_1$ accepts. The cost of predicate, $c(p)$, is the amount of the resources consumed to evaluate $p$ on a typical photo.

In this work, we focus on the smartphone energy consumption as the cost metric. Executing a query in a smartphone can be energy-hungry. For example, our measurements show that executing a face detection query on 100 photos in Nexus One costs about 300 Joules, which is 1.6% of the total battery capacity.

Figure 3 shows an example query, called Query_1, which looks for photos that contain people's faces with a large cloudy sky background. This query has three predicates. The face detection predicate finds faces in photos. Texture matching examines photos with texture similar to a cloudy sky texture. RGB thresholding only accepts the photos that have high blue color intensity to ensure the large size of the sky background. These three predicates have decreasing cost. Figure 4 shows two examples of smartphone photos accepted by this query. The patches in the figure, which contain people's faces and a cloudy sky area, show results of the face detection and texture matching predicates. The RGB thresholding predicate has favored a large sky.

### 2.2 Theia Server

Theia Server runs in powerful computers in the cloud. It has four modules: Query Distributor, Result Collector, Data Cache, and Partition Agent. Query Distributor distributes queries and maintains the state information of a query and its refinements. Result Collector gathers search results for the search user to retrieve. Data Cache stores photos offloaded from smartphones from previous searches. Because executing a query with photos in Data Cache is faster and incurs negligible cost compared to those in a smartphone, Theia always starts executing a query by using photos in Data Cache. Finally, Partition Agent works with Theia Mobile to execute offloaded search tasks from smartphones.

Theia Server enforces an incentive mechanism and a cost model on other Theia components. Theia requires such incentive mechanism and cost model in order to charge a search user for executing a query, and in order to properly motivate smartphone users to participate and to compensate them for the search energy cost and for valid search results. Theia is not tied to any particular incentive mechanism or cost model, although it assumes certain properties for them, as will be described in Section Incremental Search.

## 2.3 Theia Mobile

Theia Mobile runs inside a smartphone. It has three modules: *Search Engine, Energy Profiler,* and *Data Manager*. Search Engine receives queries from Query Distributor in Theia Server and executes them on photos. It also collaborates with Partition Agent in Theia Server to dynamically partition the execution of a query in an energy-efficient manner. Moreover, Search Engine reports identified photos along with their matching score to Result Collector in Theia Server. Energy Profiler produces the required energy measurements for Search Engine. Data Manager maintains the searchable photos in the device. It also stores the state information about previous searches for the stored photos.

## 2.4 Theia Gate

Theia Gate is where the search application is realized. It provides mechanisms for users to compose secure queries, to choose the cost budget, and to provide feedback for Theia. It also streams and visualizes search results and feedback from Theia Server as soon as some results are available.

## 3. Incremental Search for Cost Control

Searching into others' smartphones cannot be free because it consumes precious smartphone resources, e.g., battery; and because there must be an incentive for smartphone owners to participate. Instead of executing a query on all smartphones and all photos by default, Theia enables a search user to expand the search scope incrementally. A Theia user always submits a cost budget along with a query. Coupled with a cost model, the budget limits the scope of the execution, i.e., numbers of smartphones and photos, so that the search user can provide feedback or refine the query before expanding the scope.

Theia requires a cost model to charge a search user for executing a query and to compensate the smartphone users for allowing the search. The cost model is enforced by Theia Server. While Theia can support a variety of cost models, it makes two assumptions. First, Theia assumes the cost of executing a query in a smartphone consists of three parts: a flat entry cost per smartphone, a cost per searched photo, and a cost per search result. Second, Theia assumes that the cost per search result is significantly larger (by an order of magnitude) than the other two parts. This cost structure not only motivates a search user to devise a good query but also rewards smartphone users who produce interesting photos. It reflects the cost of accessing other's photos as the dominant cost.

Given the budget, Theia Server first determines *N*, the number of smartphones to search. If *N* is too large, most of the budget goes to the per-smartphone flat cost. If *N* is too small, all the results come from only a few devices, which can reduce the chance of finding relevant photos. Theia uses a simple tradeoff heuristic that allows a fixed fraction of the budget to go to the per-device flat cost.

Once the number of smartphones to search is determined, Theia must determine which smartphones to search and divide the budget equally between the selected smartphones. When a query is submitted for the first time, Theia selects smartphones randomly. When the query is submitted again, Theia gives priority to the devices from which search results have been marked by the search user as interesting. This is based on a heuristic that if the search user finds one photo from a smartphone interesting, he is more likely to find more interesting photos from the same smartphone than from a randomly selected smartphone. We refer to this property as *relevance locality*, which will be discussed further later.

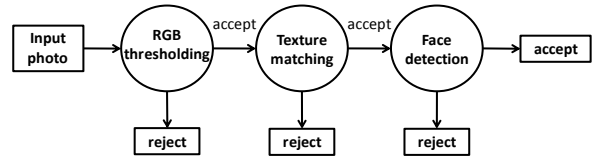

**Figure 5: Ordered execution of Query_1**

Once a smartphone is selected and the per-smartphone budget is determined, Theia first searches all the photos offloaded from the smartphone in Theia Server Data Cache and then randomly selects photos from the smartphone to search until the designated budget is reached. When randomly selecting photos, Theia skips photos that have been searched before with the same query or have been cached on Theia Server.

As the execution of a query goes on, the matched photo will be streamed to Result Collector in Theia Server along with their matching score. The matched photos will also be saved in Theia Server Data Cache to serve future searches.

The Theia Gate streams the search results from Theia Server along with information regarding the performance of the query, and visualizes the feedback. Streaming begins as soon as some results are available, even before the query execution completes. Theia provides two measurements so that the user can assess his query. The selectivity of each predicate helps user to identify over-selective and under-selective predicates. The matching scores of the returned photos for each predicate help users to refine the predicates.

## 3.1 Keeping State Information for Incremental Search

To skip searched photos when the same query is submitted again, Theia keeps state information for photos, if they have already been searched by that query. Theia identifies a query uniquely by an integer-type ID, which is generated by Theia Gate. Search Engine in Theia Mobile creates a SQLite database for each query to store the names of the photos that have been searched before or cached on Theia Server. We call this database *query state database*.

Since state information has to be looked up and stored for each photo searched, this design might seem inefficient both computation-wise and storage-wise. But it is indeed quite efficient in practice for two reasons. First, smartphone owners will mostly have less than thousands of photos in their storage. Second, since search is incremental, not all of the photos in the device will be searched by each query. Finally, most queries have short lifespan, and their state information can be discarded soon, e.g., by the end of the day.

We profiled the computation and storage overhead of such lookup and store in our implementation of Theia Mobile. Our measurements show that the lookup, which has a complexity of $O(n)$, takes less than 10ms and 30ms for database size of up to 1000, and 10000, respectively. Store, which has a complexity of $O(1)$, takes less than 30ms. Since executing the predicates in smartphones typically takes hundreds or thousands of milliseconds, we consider such overheads to be negligible. Also, databases of the mentioned sizes occupy less than 50 KB and 300 KB, respectively, which is also negligible compared to the storage capacity of smartphones.

## 4. Partitioned Search for Energy Efficiency

Executing queries in smartphones incurs high energy consumption not only because predicates can be computation-intensive but also because there can be many photos in the device to search. One obvious solution for reducing the energy cost of a compute-intensive task on mobile devices is to execute the task in the cloud, also known as offloading. However, the cloud does not have the photos in the smartphone and therefore, these files need to be uploaded too. Since there can be many photos in the device, simply offloading all of them, or *full offloading*, may not necessarily be the most energy-efficient. Therefore, we investigate the possible merits of offloading only part of the query, or *partitioned search*. With partitioned search, some of the predicates are evaluated locally in the smartphone and the rest are evaluated in Theia Server. Only the photos accepted by all of the local predicates are offloaded to Theia Server for further evaluation. In other words, only those photos that show promise are offloaded for further processing.

### 4.1 Problem Definition

Given a query, the partition problem is to identify the *order of evaluation* for the query's predicates and the *first predicate to offload* so that the total local energy cost is minimized. The order of evaluation for the predicates is important for efficiency because photos rejected by a predicate do not have to be evaluated with a later predicate [3] (Figure 5). In the case of partition, if a photo is rejected before the first offloaded predicate, the photo does not need to be offloaded to the cloud. On the other hand, if the photo is not rejected before the first offloaded predicate, it will be evaluated with the remaining predicates on Theia Server.

### 4.2 Partition Algorithm

Finding the optimal partition is not trivial. The energy cost of a partitioned search is determined by the energy cost of network activity, predicate selectivity, and predicate cost, which have to be estimated at runtime. Theia solves this problem by a two-phase solution. The *training phase* estimates the cost of predicates and wireless transfer by evaluating all the predicates on a few photos locally and offloading a few photos to Theia Server. With the cost estimations, the Search Engine determines an initial partition. The *evaluation phase* starts with the initial partition. It updates the predicate cost and estimates predicate selectivity with adaptive sampling [4] with each photo evaluated. The partition is updated after evaluating every five photos.

When creating a partition, Theia first determines the order of evaluation for all the predicates in the query and then determines the first predicate to offload. We describe these steps below.

*4.2.1 Predicate Ordering*

Theia leverages an important database concept, *conditional rank* [5]. Given the execution order of the predicates, the conditional rank of a predicate is defined as the cost of the predicate divided by one minus the selectivity of the predicate, conditioned on the predicates that come before in the order. A simple heuristic to approach the optimal execution order is to ensure that the conditional ranks of the predicates are in the same order as the predicates. This heuristic is based a key observation that if the order of execution is optimal, the conditional ranks of the predicates are in the same order. Database research [6] has shown that this heuristic

```
<?xml version="1.0" encoding="ISO-8859-1"?>
<query id="848753739">
  <and number_of_predicates="1">
    <predicate name="Face (front)" type="C">
      <arguments predicate_object="libface-predicate.so"
        init_function="f_init_opencv_fdetect"
        eval_function="f_eval_opencv_fdetect"
        fini_function="f_fini_opencv_fdetect"
        blob="haarcascade_frontalface_default.xml"
        predicate_object_version="2"/>
      <parameters num="6" p0="1.2" p1="24" p2="24"
        p3="1" p4="1" p5="4"/>
      <dependencies num="0"/>
      <threshold value="1"/>
    </predicate>
  </and>
</query>
```

**Figure 6: XML representation of a face detection query. Certain details are suppressed for clarity**

achieves a performance no worse than ~2x of the optimal solution for queries with less than 20 predicates; and it achieves the optimal in most of the cases.

Since Theia queries usually have a small number of predicates, we adopt this conditional rank-based heuristic and our experience also confirms its effectiveness. In the evaluation phase, Theia updates the cost and conditional selectivity of the predicates in their current order of execution after evaluating every photo. After evaluating every 5 photos, Theia checks to see whether enough samples are available [4] to meaningfully estimate the conditional ranks. Theia then updates the conditional ranks of those predicates for which enough samples are available, and reorders them based on their updated conditional ranks. It then discards the previous conditional ranks of the reordered predicates – since they are not valid anymore in the new order – and acquires new estimates by evaluating more photos.

*4.2.2 Partition Point*

Once the execution order of predicates is determined as above, Theia determines the partition point, or the first predicate in the order to offload, by using a special predicate, *pw* [7]. *pw* has a cost equal to the average energy cost of offloading a photo under current networking conditions, a selectivity of zero, and is independent from the predicates in the query. Therefore, the conditional rank of *pw* is always equal to the wireless transmission cost.

To find the optimal partition point, Theia simply finds the order of execution of all the predicates including *pw* using the heuristic discussed above. The predicates before *pw* are evaluated locally and those after *pw* are offloaded.

The cost of offloading a photo directly affects the partition point. Since the wireless connectivity is highly variable due to mobility, the optimal partition point can change quickly. However, since *pw* is independent from the rest of the predicates, its position can be changed without disturbing the order of execution of the query predicates. Therefore, upon detecting a change in the wireless cost, Theia can rapidly calculate the new optimal partition by merely changing the position of the wireless predicate. We call this *dynamic partition*.

## 5. Prototype Implementation

## Theia Query

We have implemented the query in two parts: the *query specification*, and the *predicate objects*. The query specification is an XML file that specifies the query ID and the predicates in the query. Figure 6 shows the XML representation of a face detection query, which has a face detection predicate only. The query specification also determines the predicate objects that must to be used for executing the predicates. For example, `libface-predicate.so` in the `<arguments>` element is the predicate object for the face detection predicate, as shown in Figure 6. The predicate objects are implemented in C or Java, as specified in the XML file in the `<predicate>` element. The C predicates are shared objects that are cross-compiled for the instruction set used in target smartphones. The Java predicates are JAR files. Android OS, which we have used in our current prototype, supports both types of predicates.

We construct three example queries that we consistently use in our experiments with Theia. Query_1 is shown in Figure 3. Query_2 is constructed from Query_1 by removing the texture matching predicate, and Query_3 is constructed from Query_2 by removing the RGB thresholding predicate, and therefore is a face detection query.

## Theia Server

We have implemented the modules of Theia Server in various programming languages and hosted it in a server on a university campus. We have implemented Query Distributor, Result Collector, and part of Data Cache in PHP and run them on an Apache web server. We also use MySQL databases in these modules to store the state information for incremental search. We have implemented Partition Agent and the other part of Data Cache in Java and run them on a Jetty web server.

Query Distributor uses two methods to send push notifications to the Search Engine in Theia Mobile. The main method is Android Cloud to Device Messaging (C2DM) [8]. We also use SMS push notification as a backup method, since we observed that C2DM fails occasionally.

To implement partitioned search, Search Engine in Theia Mobile employs a multipart HTTP request to send offloaded predicates and photos to Partition Agent, which then executes the predicates on the photo and returns the accept/reject result to the device in the HTTP response. Since the Partition Agent has access to all the predicate objects in Theia, the search engine has to enclose only the query specification in XML and a list of predicates to execute remotely (a total of few kilobytes only) in the HTTP request.

For the cost model, we use 1, 1, and 10 units for the flat cost, the cost per searched photo, and the cost per search result, respectively. These values are consistent with Theia's assumption that the cost per search result be significantly larger than the other two costs.

## Theia Mobile

We have implemented Theia Mobile for Android-based mobile systems. Search Engine can execute both C and Java predicates. It evaluates the C predicates with an executable, predicate-runner, that loads the predicate object using dynamic loading. Search Engine evaluates Java predicates using Java Reflection.

We have implemented a simple yet effective energy profiler that constructs a system energy model with linear regression based on the execution time of a predicate and wireless transfer time. Measurements show that the constructed energy model has an average error of 3% and 13% in estimating the energy cost of predicate evaluation and that of transmitting photos, respectively.

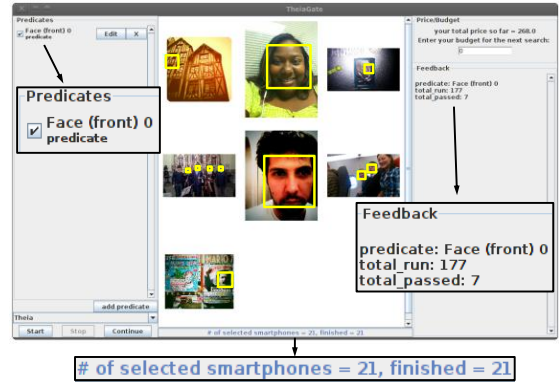

**Figure 7: A snapshot of Theia Gate in use. The search user is using a face detection query**

### 5.1 Theia Gate

We have implemented Theia Gate in Java with a graphical user interface. Theia Gate provides a set of predicate templates that the search user can leverage to generate queries. Currently, Theia Gate supports multiple predicate templates including face and body detection that use Haar feature based classifiers, texture matching, RGB thresholding, and RGB histogram matching. An RGB histogram matching predicate looks for photos that have similar RGB histogram characteristics as the input patch. Examples of the functionality of the rest of the predicates were explained in Section Theia Architecture.

To leverage the incremental search supported by Theia, Theia Gate allows a search user to assign the budget for a query. It retrieves the query performance feedback from Theia server and presents it to the search user in the end of each search. Finally, Theia Gate allows the search user to modify the predicates by changing the parameters of the templates.

Figure 7 shows a snapshot of Theia Gate. The search user is using a face detection query. He has assigned a budget in the first search, and has received 7 matching photos, all relevant except for the first one. 177 photos are searched over 21 smartphones according to the feedback on the right column, and the status bar in the bottom.

## 6. Evaluation

We evaluate Theia's effectiveness in helping search users reduce cost and in improving the energy efficiency of searching photos inside smartphones, through both user study and measurement. We also demonstrate the real-time performance of Theia using a field trial with six smartphones and photos from real smartphone users.

### 6.1 User Study of Incremental Search

We evaluate how well the incremental search feature of Theia helps search users reduce the cost of search and retrieve better results. We conduct a user study with 10 participants to use Theia Gate and perform a search task.

#### 6.1.1 Apparatus, Data Set, Participants, and Procedure

To evaluate incremental search in a large scale, we emulate 85 smartphones with Theia Mobile. Each emulated smartphone is a PHP script that can run on any PC. We implement the script so that the search speed of the emulated device is very close to that of a real smartphone with the wireless link considered. This ensures that the interactive experience with the emulated device is very close to that with a real one.

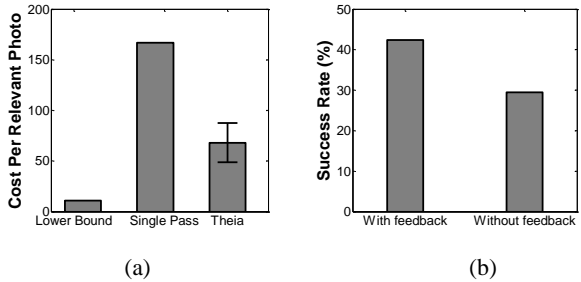
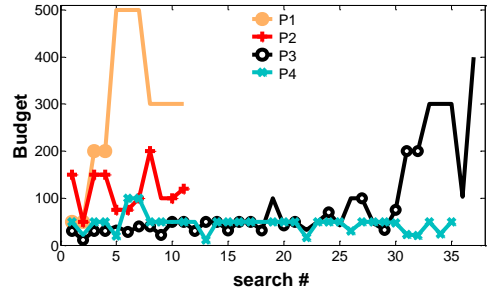

| (a) | (b) |

**Figure 8: (a)** Search cost per relevant photo (bar shows the average for Theia over all participants and error bar shows min-max), **(b)** Effectiveness of using search user feedback in selecting smartphones

**Figure 9:** Search processes for four participants: X axis indicates the order of searches performed by a participant; Y axis indicates the budget submitted with each search; A marker indicates a new or revised query is used

Each emulated smartphone is loaded with smartphone photos captured by a Flickr user. We crawled Flickr.com to collect public photos taken with smartphones including various iPhone and HTC smartphones. We collected 85 users with a total 3055 photos to emulate 85 Theia Mobiles.

We recruited 10 participants for the user study. Eight of the participants are male. All participants are students from a US private university with an average age of 24 and sciences and engineering background. We recruited the participants through flyer and direct contact, and compensated each with a $20 gift card.

The user study consisted of training, competition in a search task, and interview. We first trained a participant to use Theia Gate for about 25 minutes. We instructed them about the cost model, how to compose and revise queries, how to set the search budget, and how to provide feedbacks in Theia Gate. Then, we asked the participant to find 20 photos with cloudy sky using the emulated setup described above. To properly motivate the participants, we told them that they are in a competition with other participants based on the total search cost to find the 20 photos. The participants were allowed to set the budget and revise queries freely. After the participants found the 20 photos, they answered a survey about their experience with Theia and were interviewed further if necessary.

*6.1.2 Search Cost*
Our results show that *Theia's incremental search enables all participants to significantly reduce the cost per relevant photo*. Although the specific cost model described in Section Prototype Implementation is used for the user study, we expect the conclusion holds for all cost models in which the cost of a matched photo dominates, an assumption made by Theia's design. Figure 8(a) shows the cost per relevant photo, i.e., a photo with cloudy sky, for incremental search as achieved by the participants (Theia). Figure 8(a) also shows the cost for two hypothetical cases, *Lower Bound* and *Single Pass*. Both hypothetical cases assume a perfect query that will only return relevant photos. *Lower Bound* is the theoretical minimum cost of the same search task when all the 20 relevant photos come from a single device and only 20 photos are searched. *Single Pass* searches all the photos in all smartphones without budget constraint to return 20 relevant photos. It represents the lower bound for the cost using non-incremental search.

The results show that incremental search assists search users to effectively reduce the cost per relevant photo by an average of 59% compared to Single Pass. We expect incremental search will reduce the cost even more significantly in real deployments where there are more smartphones and photos to search. On the other hand, the cost of incremental search is on average 6 times larger than the theoretical minimum, which shows that there is still substantial room for improvement in our implementation.

Our results further show that *the search user's feedback also helps*. Figure 8(b) shows the success rates of search into devices from which search results are and are not marked by the participants as relevant in the previous searches, respectively. The success rate is defined as the number of relevant photos divided by the number of search results. We see that Theia's use of the user feedback increases the success rate by 44% compared to searching the smartphones that are not marked by the search user.

*6.1.3 Participants' Interaction with Theia*
By monitoring the participants, we are able to inspect their interaction with Theia. Figure 9 shows the search processes by four participants, P1 to P4. P2 and P4 incurred the lowest cost among the 10 participants; and P1 and P3 the highest. The X axis denotes each search (or submission of a query) in the order of performance and the Y axis denotes the budget the participant chose for each search. A marker indicates the participant submitted a new query, usually a revised one. The number of searches and that of the revisions collectively indicate how much time and effort a user spends.

We make the following observations. First, the 10 participants used Theia in very different ways, leading to a large range of total cost (from 973 to 1753 units), a large range of number of searches (from 9 to 37) and a large range of number of revisions (from 1 to 31). Second, while a few participants like P2 finished the search with low cost and a small number of searches and revisions, most participants made a tradeoff between cost and the effort. For example, P4 used small budgets and revised a lot to reduce the total cost, while P1 used large budgets and finished with much fewer revisions and searches. Finally, a moderate budget 10 to 20 times of the cost per search result seems to work well as used by P2 and several other participants. A budget too small as used by P3 and P4 will lead to more searches not only because a very small budget will pay a few results but also because the search user receives less feedback from Theia and can provide feedback only for a few results to help future searches. On the other hand, a budget too large as used by P1 can be wasteful, in particular when the query is not well refined yet. Since our participants only received training of 25 minutes, the above observations strongly suggest more training and experience will help Theia users significantly improve their productivity.

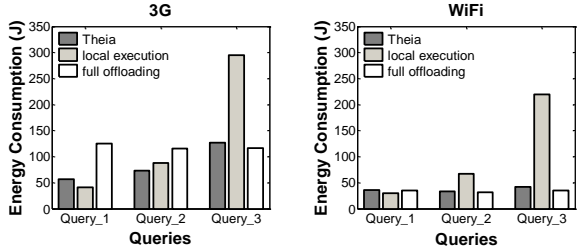

Figure 10: Total smartphone energy consumption of searching 100 photos with 3G and WiFi connectivity

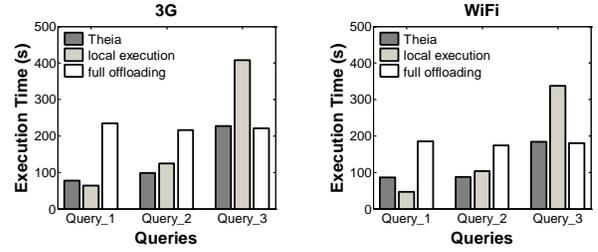

Figure 11: Execution time of searching 100 photos in a smartphone with 3G and WiFi connectivity

All but two of the participants (P1 and P3) found it easy to learn the concepts of Theia and work with it. P1 and P3, not surprisingly, were frustrated by the large total cost and, in P3's case, a large cost despite a lot of effort. All participants would like to have more predicate options to compose and revise queries. Therefore, enhancing Theia Gate for richer and more flexible queries is our immediate future work on this project.

## 6.2 Measurement of Partitioned Search

We conduct controlled experiments to evaluate the effectiveness of partitioned search in improving energy efficiency. We execute the three example queries, Query_1, Query_2, and Query_3, in a Nexus One smartphone with around 100 photos from a real user's smartphone. For each example query, we measure the energy consumption of the Nexus One when using partitioned search. We repeat all the measurements when the device uses local query execution and full offloading. Moreover, to evaluate the partitioned search in different network conditions, we repeat each experiment for both the WiFi and the 3G connection. The WiFi connection has an average power draw of 266 mW for transmission, and shows median RTT of 66 ms between the smartphone and Theia Server, which are 1140 miles apart. The 3G connection has an average power draw of 571 mW for transmission, and shows median RTT of 95ms.

The results, summarized in Figure 10, show that *partitioned search reduces the energy consumption of executing the search by up to 55% and 81% compared to full offloading and local execution, respectively.* More importantly, partitioned search improves the efficiency without slowing down the search. As shown in Figure 11, partitioned search reduces the query execution time significantly compared to full offloading and local execution in most of the experiments.

To evaluate if partitioned search adapts to changes in the wireless link well, we repeat the experiment with Query_1 using the WiFi network with a one second delay injected into the network connection in the middle of the experiment. Figure 12 illustrates the partitioning of predicates of Query_1 throughout the experiment. It demonstrates that the partitioned search algorithm detects the change in the wireless connection rapidly (after evaluation of a few photo), and adapts to the new condition by executing the texture matching predicate locally.

## 6.3 Field Study

We conduct a field study to assess the real-life experience with Theia. Our testbed consists of six Android smartphones with Theia Mobile installed. In particular, we are interested in how fast search results can be retrieved considering the distributed, wireless, and resource-limited nature of smartphones. The smartphones include three HTC Nexus One's, two Motorola Droids, and one Samsung Galaxy S. One of HTC Nexus One's use T-Mobile 3G network,

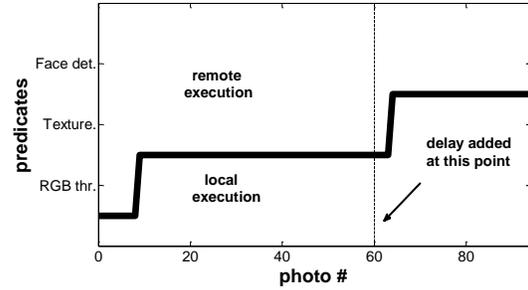

Figure 12: Theia adapts to network condition change through dynamic partition. X axis shows the order of photos evaluated by Query_1; Y axis shows the predicates in Query_1; The thick line shows the border between the predicates that are executed locally and remotely

one of Motorola Droids use Verizon 3G network, the Samsung Galaxy S uses AT&T 3G network, and the rest use a university WiFi network. The smartphones are in a different USA state from where Theia Server is hosted or 1140 miles apart.

Each smartphone is loaded with photos collected from the smartphone of a real user. We collected photos from the smartphones of 11 participants. This allows us to repeat each experiment with photos from two different participants. The participants are all undergraduate students from a private university in the USA. The average number of photos we collected from each participant is 189, another evidence that smartphone users leave a lot of photos in their devices.

We conduct two sets of experiment, and for each set, we choose photos of 6 participants and store them in the phones (with one overlap between two sets). We then submit three queries from Theia Gate, *All_Accept*, Query_2, and Query_3. All_Accept is a special query that accepts all the photos it searches without any processing. It represents a lower bound on the latency of result retrieval. Compared to All_Accept, Query_2 (similar to Query_1) and Query_3 have much lower selectivity and much higher execution time, respectively, which slow down result retrieval.

First, we investigate the latency of retrieving the first search result from the testbed, as shown in Figure 13(a). The results show that the latency of retrieving the first result is as low as 4 seconds in All_Accept and no more than 30 seconds in Query_2 and Query_3.

Second, we investigate the time interval between retrieving the consecutive results from the testbed, as shown in Figure 13(b). The results show that the median interval between consecutive results is as low as 0.7 seconds in All_Accept and is no more than 7 seconds in Query_2 and Query_3.

Figure 13 also shows the latency of retrieving the first result and the interval between consecutive results for a single smartphone. We see that the latency increases noticeably with only one smartphone. These results show that increasing the number of smartphones reduces the latency of result retrieval significantly in Theia. Therefore, we expect that latency in Theia will be further reduced in real deployments with many more smartphones.

We also found that it takes a median of 5 seconds for each device to receive the search push notification from Theia Server.

## 7. Related Work

To the best of the authors' knowledge, Theia is the first search system that treats resource-constrained smartphones as real-time searchable photo databases. No existing photo search system supports incremental and partitioned search, which are the key to Theia's capability to control search cost and improve search efficiency. While prior work has studied distributed, resource-constrained sensor nodes as databases, e.g., TinyDB [9], search in such databases is predefined and the retrieval of search results through multiple network hops incurs most of the energy cost. In contrast, search is opportunistic in Theia and the execution of query inside the database (smartphone) incurs most of the energy cost due to the compute-intensive nature of photo content search. As a result, Theia faces a very unique set of technical challenges.

All existing photo search systems such as images.google and Diamond [3] host databases in powerful servers. They focus on making search results relevant and returned fast. There is no need for incremental or partitioned search. Moreover, images.google indexes photos and supports textual queries. In contrast, indexing photos would be impractical to opportunistic search in Theia since the queries are not known a priori.

Theia's query design draws upon results from research in relational databases [6, 10, 11]. However, unlike queries in relational databases that are textual, queries in Theia are XML data structures and photo-processing code objects. Partitioned search in Theia leverages ideas in query optimization in relational databases [6, 10, 11]. However, instead of minimizing the query execution time in a server-hosted database, Theia's partitioned search minimizes the energy consumption of query execution.

There is a wealth of research on task offloading and remote execution for mobile devices in order to leverage the resources in the cloud and save resources in the device, e.g., [12]. Unlike existing work that target offloading for a program with a known order of execution, partitioned search is designed for ordering and partitioning predicates that have no pre-determined order of execution.

The fundamental motivation of Theia is similar to that of participatory sensing applications [13-16]. That is, data captured by a smartphone user may be useful to others. However, Theia differs from participatory sensing in how data captured by a smartphone user is made useful to others. While smartphone users share pre-determined data in participatory sensing applications, they do not know which photos to share in Theia. As a result, Theia is realized as a search system rather than a sensor network.

## 8. Discussions and Future work

While this paper focuses on the system design and evaluation of Theia, we next discuss several important issues that we plan to address in the future.

### 8.1 Privacy and Security

Similar to participatory sensing applications, protecting smartphone owners' privacy is vital for wide adoption of Theia. A simple solution is to ask smartphone users who participate to tag photos for Theia search or simply store them in a special folder, and Theia Mobile's Search Engine will only examine these photos. The current Theia prototype adopts this solution. Interviews with the participants in our user study suggest that such a simple arrangement is indeed usable and acceptable because it is mentally similar to how people share photos on-line already. However, more sophisticated solutions to simplify user's effort in protecting privacy may be needed for real-world deployment.

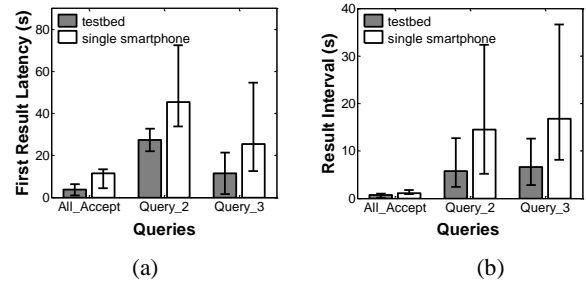

**Figure 13: (a) Latency of getting the first result, (b) Interval between consecutive results. Bars show the median and error bars show 25 and 75 percentiles**

The opportunistic nature of Theia also invites a security concern because a Theia query is a piece of code created by a search user to execute inside others' smartphones. Since Theia Gate only allows search users to compose and revise queries with given predicates and their parameters, our current prototype dodges this concern. On the other hand, the architecture of Theia does provide several means to address the security concern in a more rigorous manner. First, Theia Mobile's Search Engine can sandbox query execution using well-known techniques [17]. Moreover, Theia Server can leverage its computational power to verify and test queries with automatic software test technologies similar to that provided by [18].

### 8.2 Relevance Locality

A key feature of Theia is to allow search users to mark search results that they find relevant. When the same query is submitted again, Theia will give a higher priority to smartphones from which the relevant photos are retrieved. The evaluation showed this feature helps the effectiveness of search significantly.

The effectiveness of this simple feature suggests something significant: *relevance locality*. That is, relevant results are very likely to come from the same database (smartphone in our case) and maybe also from similar databases. This is not surprising in view of the temporal and spatial locality of smartphones and the relatively stable personal interest of a smartphone user. For example, if a photo with the lost child in our example is found from a smartphone, it is likely more relevant photos may be in the same smartphone and smartphones that have taken photos from a similar location and time. Such relevance locality can be true to any distributed database that stores acquired data locally, including smartphones and wireless sensor nodes.

While Theia already capitalizes relevance locality in smartphone photos by simply treating smartphones with relevant photos favorably, we plan to further study relevance locality to improve the scoping of opportunistic search.

## 9. Conclusion

We reported the first working system that allows content-based search of photos inside smartphones. By using incremental search,

Theia helps search users to effectively reduce the cost per relevant photo. The use of user's feedback to refine search scope also helps to retrieve more relevant photos, thanks to relevance locality. By using partitioned search, Theia reduces the energy consumption of executing the search, even under changing network conditions. Theia returns results with median latency of seconds from a single smartphone. Finally, Theia is an important first step toward opportunistic content search of smartphone photos. It invites further research into many interesting problems when users search smartphones for photos that interest them.